\documentclass [aps,prb,superscriptaddress,twocolumn,showpacs,a4paper,amsmath,
amssymb,byrevtex,floatfix]{revtex4}

\usepackage[latin1]{inputenc}
\usepackage{graphicx}
\usepackage{dcolumn}
\usepackage{bm}
\usepackage{hyperref}
\usepackage{times}
\usepackage{multirow}

\begin{document}

\title{Symmetry-induced interference effects in metalloporphyrin wires}

\author{R. Ferrad\'as}
\affiliation{Departamento de F\'{\i}sica, Universidad de Oviedo, 33007 Oviedo Spain}
\affiliation{Nanomaterials and Nanotechnology Research Center, CSIC - Universidad de Oviedo, Spain}

\author{V. M. Garc\'{\i}a-Su\'arez}\email{vm.garcia@cinn.es}
\affiliation{Departamento de F\'{\i}sica, Universidad de Oviedo, 33007 Oviedo Spain}
\affiliation{Nanomaterials and Nanotechnology Research Center, CSIC - Universidad de Oviedo, Spain}
\affiliation{Department of Physics, Lancaster University, Lancaster LA1 4YW, United Kingdom}

\author{J. Ferrer}
\affiliation{Departamento de F\'{\i}sica, Universidad de Oviedo, 33007 Oviedo Spain}
\affiliation{Nanomaterials and Nanotechnology Research Center, CSIC - Universidad de Oviedo, Spain}
\affiliation{Department of Physics, Lancaster University, Lancaster LA1 4YW, United Kingdom}

\date{\today}

\begin{abstract}
Organo-metallic molecular structures where a single metallic atom is embedded in the organic backbone are ideal systems to study the effect of strong correlations on
their electronic structure. In this work we calculate the electronic and transport properties of a series of metalloporphyrin molecules sandwiched by gold electrodes
using a combination of density functional theory and scattering theory. The impact of strong correlations at the central metallic atom is gauged by comparing our
results obtained using conventional DFT and DFT$+U$ approaches.
The zero bias transport properties may or may not show spin-filtering behavior, depending on the nature of the $d$ state closest to the Fermi energy. The type of
$d$ state depends on the metallic atom and gives rise to interference effects that produce different Fano features. The inclusion of the $U$ term opens a
gap between the $d$ states and changes qualitatively the conductance and spin-filtering behavior in some of the molecules. We explain the origin of the
quantum interference effects found as due to the symmetry-dependent coupling between the $d$ states and other molecular orbitals and propose the use of these systems
as nanoscale chemical sensors. We also demonstrate that an adequate treatment of strong correlations is really necessary to correctly describe the transport
properties of metalloporphyrins and similar molecular magnets.
\end{abstract}


\maketitle

\section{Introduction}
A key issue in the field of molecular electronics is the search for molecular compounds that give rise to new or improved functionalities.
Porphyrin molecules constitute in this respect promising candidates and are as such receiving increased attention. These molecules play an essential role in many biological
processes such as electron transfer, oxygen transport, photosynthetic processes and catalytic substrate oxidation. \cite{Dolphin78}. Porphyrins have been extensively studied
in the past by biologists and chemists \cite{Dorough51,Gust76,Goff76,D'Souza94}. However, an increasing number of theoretical \cite{MengSheng02,Palummo09,Rovira97} and
experimental \cite{OtsukiSTM10,VictorNature11} physics analyzes have appeared in the past few years. Progress in the design of supramolecular structures involving
porphyrin molecules has been rather spectacular \cite{Irina09}.

Porphyrins are cyclic conjugate molecules. Their parent form is the porphin, which is made of four pyrrole groups joined by carbon bridges and has a nearly planar D4h
symmetry \cite{Jentzen96}. The alternating single and double bonds in its structure, make this molecule chemically very stable. In porphyrin systems, the porphine is
the base molecule, but different functional groups can join the macrocycle by replacing a peripheral hydrogen. In addition, the macrocycle can accommodate inside a metallic
atom, such as Fe (which is the base of the hemoglobin in mammals), Cu (hemolymph in invertebrates), Mg (chlorophyll in plants), etc. Large porphyrin systems can undergo
certain ruffling distortions because of its peripheral ligands, its metallic atom inside or the environment \cite{PaikSuh84}. It is precisely the large number of possible
configurations, which give rise to a wide variety of interesting properties that make these molecules very attractive for molecular-scale technological applications.

\begin{figure}
\includegraphics[width=\columnwidth]{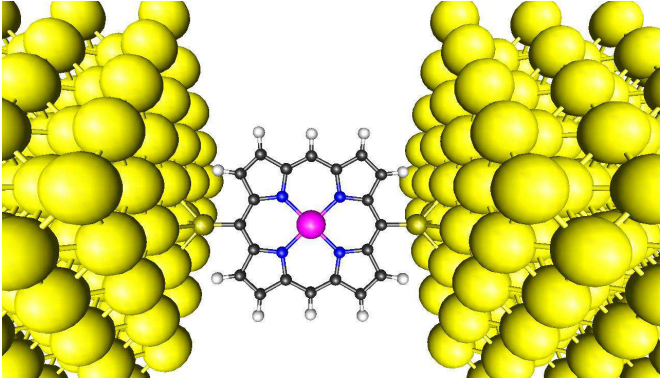}
\caption{\label{AuPhFe}(Color online) Schematic view of a Fe-porphyrine molecule between gold leads. Yellow, dark yellow, grey, black, blue, and magenta represent gold, sulphur, hydrogen, carbon, nitrogen and iron, respectively.}
\end{figure}

In this work, we present a theoretical study of the electronic and transport properties of porphyrins sandwiched by gold electrodes using Density Functional Theory (DFT).
The porphyrine molecule is attached to the gold surface by a thiol group and oriented perpendicular to it, as sketched in Fig. (\ref{AuPhFe}). The metallic atom placed
inside can be Fe, Co, Ni, Cu and Zn. For the sake of comparison, we have also studied the porphine compound, which has no metallic atom. Since DFT fails to describe correctly transition metal
elements in correlated configurations, we have also adopted a DFT$+U$ approach. We show that the inclusion of strong correlations is necessary to accurately characterize the transport properties of some of the metalloporphyrin complexes.

The outline of the article is as follows: in section II we briefly explain the DFT$+U$
flavor used in our calculations. In section III we give the computational details and in the last two sections, IV and V, we present and discuss the electronic and transport
properties of the junctions, respectively.

\section{Details of the DFT$+U$ approach}

DFT has emerged as the tool of choice for the simulation of a wide array of materials and nanostructures. However, the theory fails to describe strongly correlated electron
systems, such as embedded 3d transition metal or 4f rare earth elements. Apart from the fact that even exact DFT can not describe all excited states\cite{Carrascal12},
the approximations included in
the exchange-correlation potential induce qualitative errors in correlated systems. There have been many attempts to fix these problems. These include
generating exchange-correlation functionals specifically tailored to the system under investigation, but then those are frequently not transferable.
Another attempt is based on removing the electronic self interactions introduced by the approximate treatment of exchange in DFT\cite{PerdewZunger81}. This unphysical
self-interaction is a significant source of errors when electrons are localized. However it is not clear that just by removing the self-interaction error the physical
description will be qualitatively correct. We will use in this article the popular DFT$+U$ approach, developed by Anisimov and
co-workers \cite{AnisimovZaanen91,SolovyevDederichs94,LiechtensteinAnisimov95}, that represents a simple mean-field way to correct for strong correlations in systems with
transition metals or rare earth compounds.

The DFT$+U$ method assumes that electrons can be split into two subsystems: delocalized electrons, that can be treated with traditional DFT, and localized electrons
(3d or 4f), which must be handled using a generalized Hubbard model hamiltonian with orbital-dependent local electron-electron interactions. The DFT$+U$ functional
is defined then as:

\begin{eqnarray}
E^{\mathrm{DFT}+U} [\rho^\sigma (\overrightarrow{r}) , \{ n^\sigma \}]=E^\mathrm{DFT} [\rho^\sigma (\overrightarrow{r})]+\nonumber
\\
+E^\mathrm{Hub} [\{ n^\sigma \}]-E^\mathrm{DC} [\{ n^\sigma \}]
\end{eqnarray}

\noindent where $E^\mathrm{DFT}$ is the standard DFT functional;  $\rho^\sigma (\overrightarrow{r})$ is the charge density for the $\sigma$ spin; $E^\mathrm{Hub}$
is the on-site coulomb correction; $E^\mathrm{DC}$ is the double counting correction, which is necessary to avoid including again the average electron-electron
interaction that is included in $E^\mathrm{DFT}$; and $\{ n^\sigma \}$ are the atomic orbital occupations corresponding to the orbitals that need to be corrected.

The generalized Hubbard Hamiltonian is written as

\begin{equation}
\hat{H}_\mathrm{int} = \frac{U}{2}  \sum_{m,m',\sigma}\hat{n}_{m,\sigma}\hat{n}_{m',-\sigma}+\frac{U-J}{2}  \sum_{m\neq m',\sigma}\hat{n}_{m,\sigma}\hat{n}_{m',\sigma}
\end{equation}

\noindent Following Ref. \onlinecite{Dudarev98}, we take the atomic limit of the above hamiltonian where the number of strongly correlated electrons
$N_{\sigma} = \sum_{m} n_{m,\sigma}$ is an integer and write:

\begin{eqnarray}
E^\mathrm{DC}& =& \langle \mathrm{integer}\, N_{\sigma} |\hat{H}_\mathrm{int}| \mathrm{integer}\, N_{\sigma} \rangle \nonumber
\\
&=& \frac{U}{2}  \sum_{\sigma} N_{\sigma} N_{-\sigma}+\frac{U-J}{2}  \sum_{\sigma} N_{\sigma}(N_{\sigma} -1)
\end{eqnarray}

\noindent In contrast, for a noninteger occupation number, corresponding to an ion embedded in a larger system, we write:

\begin{eqnarray}
E^\mathrm{Hub} &=& \langle \mathrm{noninteger}\, N_{\sigma} |\hat{H}_\mathrm{int}| \mathrm{noninteger}\, N_{\sigma} \rangle
\\
&=& \frac{U}{2}  \sum_{m,m',\sigma} n_{m,\sigma} n_{m',-\sigma}+
\frac{U-J}{2}  \sum_{m\neq m',\sigma}n_{m,\sigma}n_{m',\sigma}\nonumber
\end{eqnarray}

\noindent The above two equations can be merged and after some algebra the DFT$+U$ functional can be written as

\begin{eqnarray}
E^{\mathrm{DFT}+U}= E^\mathrm{DFT}  + \frac{U_\mathrm{eff}}{2} &\sum_{m,\sigma} n_{m,\sigma}(1- n_{m,\sigma})
\end{eqnarray}

\noindent where $U_\mathrm{eff} = U-J$.

It has to be noted that the correction term depends on the occupation number matrix. This occupation number matrix, a centrepiece of the DFT$+U$ approach,
is not well defined, because the total density charge can not be broken down into simple atomic contributions. Since the appearance of the DFT$+U$ approach,
there have been many different definitions of the occupation number matrix \cite{CTablero08}.
We evaluate the occupation number matrix by introducing projection operators in the following way\cite{CTablero08}:

\begin{equation}
{n_{m}}^{(\sigma)} = \sum_{j} {q_{j}}^{(\sigma)} \langle {\varphi_{j}}^{(\sigma)} |{\hat{P}_{m}}^{(\sigma)}|{\varphi_{j}}^{(\sigma)}\rangle
\end{equation}

\noindent where ${\varphi_{j}}^{(\sigma)}$ are the KS eigenvectors for the $j$ state with spin index $\sigma$ and $q_{j}^\sigma$ are their occupations.
The choice of the projection operators $\hat{P}_{m}^{(\sigma)}$ is crucial, because due to the non-orthogonal basis set different projection operators give different occupation
number matrices. In our case, the choice corresponds to so-called {\em full} representation. Here, the selected operator is

\begin{equation}
{\hat{P}_{m}}^{(\sigma)} = |\chi_{m} \rangle \langle \chi_{m}|
\end{equation}

\noindent where $|\chi_{m} \rangle$ are the atomic orbitals of the strongly correlated electrons. Introducing this projector in (6), we get:

\begin{equation}
{{\bf n}_{\sigma}}^\mathrm{full} = {\bf S}{\bf D}_{\sigma}{\bf S}
\end{equation}

\noindent where ${\bf S}$ is the overlap matrix and ${\bf D}_{\sigma}$ is the density matrix of the system.

\section{Computational Method}

We have performed the electronic structure calculations using the DFT code SIESTA\cite{SIESTA}, which uses norm conserving pseudopotentials and a basis set of
pseudoatomic orbitals to span the valence states. For the exchange and correlation potential, we have used both the local density approximation (LDA),
as parameterized by Ceperley and Adler\cite{CA}, and the generalized gradient approximation(GGA), as parameterized by Perdew, Burke and Ernzernhof\cite{PBE}.
SIESTA parameterizes the pseudopotentials according to the Troullier and Martins\cite{Tro91} prescription and factorizes them following Kleynman and
Bylander\cite{Kle82}. We included non-linear core correction in the transition metal pseudopotentials to correctly take into account the overlap between the valence
and the core states \cite{Lou82}. We also used small non-linear core corrections in all the other elements to get rid of the small peak in the pseudopotential close to
the nucleus when using the GGA approximation.

We placed explicitly the $s$ and $d$ orbitals of gold as valence orbitals and employed a single-${\zeta}$ basis (SZ) to describe them. We used a double-${\zeta}$ polarized
basis (DZP) for all the other elements (H, C, O, N, S and TM). We computed the density, Hamiltonian, overlap and the matrix elements in a real space grid defined with an
energy cutoff of 400 Ry. We used a single $k$-point when performing the structural relaxations and transport calculations, which is enough to relax the coordinates and
correctly compute the transmission around the Fermi level in the case of gold electrodes. We relaxed the coordinates until all forces were smaller than 0.01 eV/\AA. We varied
the $U$- parameter and the radii of the projectors of the $U$-projectors and compared the results with experiments or previous simulations of the isolated molecules.
In this way we found that the optimal values for them were 4.5 eV and 5.5 \AA, respectively.

Fig. (\ref{AuPhFe}) shows the central part of the extended molecule in a gold-Fe-Porphirine-gold junction. The gold electrodes were grown in the (001) direction, which we took
as the $z$ axis. The sulphur atoms were contacted to the gold surfaces in the hollow position at a distance of 1.8 \AA. We carried out a study of the most stable position and
distance and found, both in GGA and LDA, that the hollow configuration was more stable than the top and bottom, in agreement with previous results obtained with other molecules.
The most stable distances were 1.6 \AA/ and 1.8\AA/ for LDA and GGA, respectively. We finally chose a distance of 1.8 \AA for all cases in order to make a systematic study of
geometrically identical systems.

We performed the transport calculations using GOLLUM, a newly developed and efficient code\cite{Gollum}. The junctions were divided in three parts: left and right leads
and extended molecule (EM). This EM block included the central part of the junction (molecule attached to the gold surfaces) and also some layers of the gold leads to make
sure that the electronic structure was converged to the bulk electronic structure. The same general parameters as in the bulk simulation (real-space grid,
perpendicular $k$-points, temperature, etc.) and also the same parameters for the gold electrodes (bulk coordinates along $x$ and $y$ and basis set) were used in the
EM simulation.
\begin{table}
\caption{Energy gaps of metalloporphyrins in the gas phase, calculated with LDA and GGA and with or without $U$.}\label{Tab01}
\begin{ruledtabular}
\begin{tabular}{lcccc}
&LDA &LDA$+U$&GGA&GGA$+U$ \\
\hline
FeP&0.10&0.85&0.48&1.10\\
CoP&0.27&1.80&0.90&1.80\\
NiP&1.55&1.65&1.50&1.60\\
CuP&0.85&1.40&0.75&1.40\\
ZnP&1.60&1.60&1.60&1.60\\
\end{tabular}
\end{ruledtabular}
\end{table}

\section{Electronic structure of metallo-porphyrins in vacuum}

Firstly we studied the effect of the $U$ term on the electronic structure of the metallo-porphyrine MP molecules in vacuum. From these initial simulations it is already
possible to fetch an idea of the main effects of correlations and their future impact on the transport properties. We chose for these analyzes a cubic unit cell of size
$35 \times 15 \times 35 \ {\AA}$ to ensure that the molecules did not interact with adjacent images.

In order to determine the influence of different $U$'s and cutoff radii and see how the results compare to previous experiments and calculations we studied
first the case of the bulk oxide FeO \cite{Parmigiani99}, where DFT is known to give qualitatively different results (metallic instead of insulating
character) \cite{Cococcioni05}. We performed calculations with $U=4$ eV and $U=4.5$ eV, which is the range of values most used in the literature for
iron \cite{Cococcioni05}, and we used projectors with different radii. We found that the results were very sensitive to these radii. Specifically, the gap
for FeO only appeared for radii larger than $2.5$ Bohr. The parameters that best fitted the experiments and previous calculations for FeO were $U=4.5$ eV
and $r_\mathrm{c} = 5.5$ Bohr, which gave a gap of $2.5$ eV (2.4 eV in Ref. \onlinecite{Parmigiani99}).

To reproduce previous theoretical results published for the iron porphyrine\cite{MengSheng02,Panchmatia08}, we had to choose $r_\mathrm{cut} = 1.5$ Bohr,
which is smaller than the values used for FeO. This is possibly justified by the fact that the values of $U$ and the projectors radii depend on the environment
where the strongly correlated atom is located. In case of the other transition metals (Co, Ni, Cu and Zn), $U$ is expected to increase towards the Zn but still
be similar \cite{SolovyevDederichs94}. We therefore used the same parameters for all metalloporphyrins, which also simplified the comparison between different
cases and made the study more systematic.

We summarize in the following subsections the main features of the electronic structure of all molecules. In some cases we show the density of states projected (PDOS)
onto the $d$ orbitals and/or the carbon and nitrogen atoms to determine the properties of the $d$ states and their relation to the other molecular states. The gaps and
magnetic moments of each molecule calculated with LDA and GGA and with or without $U$ are summarized in Tables \ref{Tab01} and \ref{Tab02}, respectively.

\begin{table}
\caption{Magnetic moments in units of $\mu_\mathrm{B}$ of metalloporphyrins in the gas phase and between gold leads, calculated with LDA and GGA and with or without $U$.}\label{Tab02}
\begin{ruledtabular}
\begin{tabular}{lcccc}
&LDA &LDA$+U$&GGA&GGA$+U$ \\
\hline
FeP&2.00&2.00&2.00&2.00\\
CoP&1.00&1.00&1.00&1.00\\
NiP&0.00&2.00&0.00&2.00\\
CuP&1.00&1.00&1.00&1.00\\
ZnP&0.00&0.00&0.00&0.00\\
\hline
AuFePAu&1.14&1.32&1.07&1.34\\
AuCoPAu&0.62&1.07&0.83&1.08\\
AuNiPAu&0.00&0.00&0.00&0.00\\
AuCuPAu&0.84&0.97&0.80&0.98\\
AuZnPAu&0.02&0.00&0.02&0.00\\
\end{tabular}
\end{ruledtabular}
\end{table}

\subsection{Iron porphyrine FeP}

\begin{figure}
\includegraphics[width=\columnwidth]{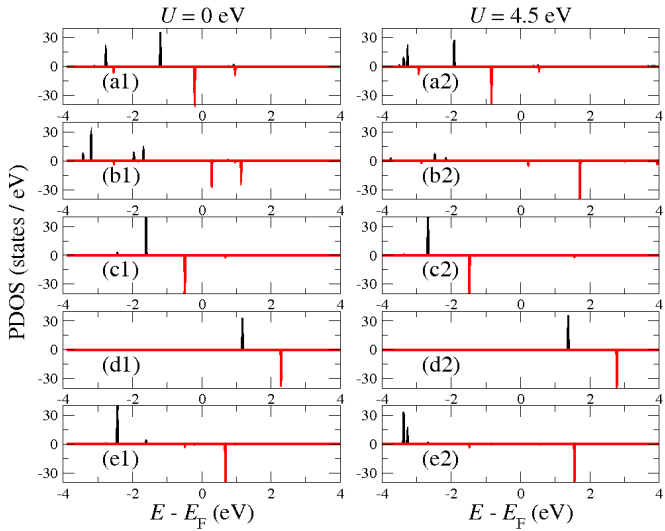}
\caption{\label{PDOS.GGA.SP.Fe}(Color online) Projected density of states (PDOS) on the iron $d$ states $- d_{xy}$ (a), $d_{yz}$ (b), $d_{3z^2-r^2}$ (c), $d_{xz}$ (d) and
$d_{{x^2}-{y^2}}$ (e) $-$ for a iron metalloporphyrin in the gas phase, computed with GGA and spin polarization. Left (1) and right (2) columns correspond to calculations with
$U=0$ eV and $U=4.5$ eV. Positive and negative values represent spin up and spin down electrons.}
\end{figure}

The lowest energy electronic configuration for FeP in all calculated cases (LDA, LDA$+U$, GGA, and GGA$+U$) is [...]$ {(d_{xy})^2} {(d_{z^2})^2} {(d_{yz})^1}
{{(d_{{x^2}-{y^2}}})^1} $. This configuration is a consequence of the crystalline field generated by the molecule, which is located on the $xz$ plane -the largest interaction
between the nitrogens and the $d$ states corresponds to the $d_{xz}$, which moves up in energy and empties-. The total spin is therefore $S_z=1$, an intermediate-spin
configuration, which is in agreement with experimental results\cite{Spin-citation}. The PDOS on the iron $d$ states calculated with GGA and GGA$+U$
is shown in Fig. (\ref{PDOS.GGA.SP.Fe}) for spin up and spin down electrons. As can be seen the closest orbitals to the Fermi level are the spin down $d_{xy}$ and $d_{yz}$.
The states that are filled and contribute to the magnetic moment of the molecule are the $d_{yz}$ and the $d_{x^2-y^2}$, whereas the $d_{xz}$ is completely empty.
The $d_{yz}$ shows also a strong hybridization. When the $U$ is included all gaps increase as a consequence of the movement of the filled orbitals to lower energies and
the empty orbitals to higher energies. From here it is already possible to get an idea on the effect of the iron states on the transport properties of the molecule. States
with large hybridization with other molecular states, such as the $d_{yz}$ are expected to give rise to relatively broad transmission resonances, whereas those states with
small hybridization such as the ($d_{z^2}$, and the $d_{x^2-y^2}$) are expected to produce either very thin resonances or Fano resonances, as explained below.

\begin{figure}
\includegraphics[width=\columnwidth]{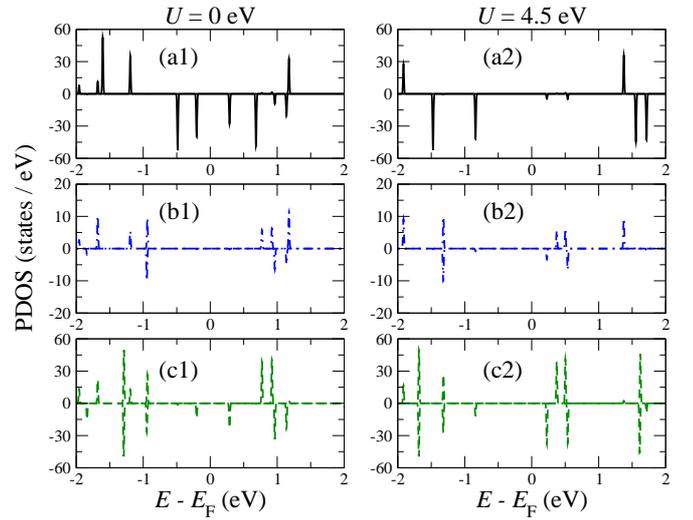}
\caption{\label{PDOS.GGA.SP.FeP}(Color online) Projected density of states (PDOS) on the iron (a), nitrogen (b) and carbon (c) states for a iron metalloporphyrin in the gas
phase, computed with GGA and spin polarization. Left (1) and right (2) columns correspond to calculations with $U=0$ eV and $U=4.5$ eV. Positive and negative values represent
spin up and spin down electrons. Notice the vertical scale is different in the middle panels (nitrogen).}
\end{figure}

The amount of hybridization can also be seen by plotting the PDOS on each type of molecular atom, as shown in Fig. (\ref{PDOS.GGA.SP.FeP}) for GGA. Notice there are iron
states that do not hybridize with the rest of the molecule, whereas other iron states do hybridize producing extended molecular orbitals. Focusing on particular states
it is possible to see that the HOMO and LUMO have Fe and C contributions. In the HOMO, the contribution of Fe is bigger but in the LUMO both Fe and C contribute equally.
Specifically, the Fe contribution to the HOMO (LUMO) comes from $3d_{xy}$ and ($3d_{yz}$) orbitals; in addition, iron contributes to the HOMO-1 with the $3d_{z^2}$ and a small
part of the $3d_{{x^2}-{y^2}}$. In the case of carbon the state that contributes to both the HOMO and LUMO is the $2p_{y}$. Note that there are also sulphur states around
the Fermi level, specially on the LUMO and below the HOMO but in order to simplify the description we focus only on the carbon, nitrogen and metallic states.
In this case the HOMO-LUMO gap is about 0.5 eV. With GGA$+U$ the states that most contribute to HOMO are still Fe $3d_{xy}$, but the C $2p_{y}$ states are the most
important in the LUMO; the LUMO has also a small contribution from N $2p_{y}$. Even though the $U$ term acts only on the iron atom, the C $2p_{y}$ states are indirectly
affected, i.e. the C $2p_{y}$ states, located around -1 eV and 1 eV, lower their energy as a consequence of the changes on the iron states. In this case the HOMO-LUMO
gap increases to 1.1 eV.

With LDA and LDA$+U$ the results are slightly different. The states that contribute most to the HOMO and LUMO are Fe $3d_{z^2}$ and C $2p_{y}$. The HOMO-LUMO gap without $U$
is about 0.1 eV. With LDA$+U$ the Fe $3d_{z^2}$ states contributes much more to the HOMO and the C $2p$ to the LUMO, and the C $2p$ states are less affected by the $U$.
The gap with $U$ increases to 0.8 eV.

\begin{figure}
\includegraphics[width=0.8\columnwidth]{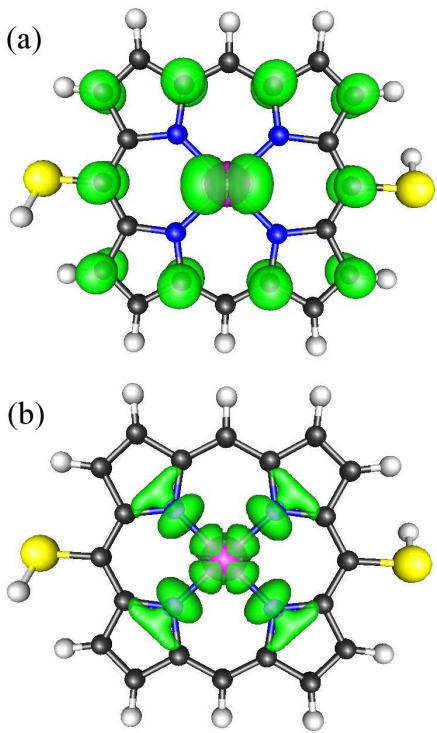}
\caption{\label{Fe.XZ-YZ}(Color online) Spatial distribution of the density of states projected on the spin-down main peak of the iron $d_{yz}$ (a) and
$d_{xz}$ (b), computed with GGA.}
\end{figure}

By using the spatial projection of the density of states (local density of states, LDOS) it is also possible to understand where a particular molecular state is located
on the molecule. In Fig. (\ref{Fe.XZ-YZ}) we show the LDOS projected on the molecular states associated to the $d_{yz}$ and $d_{xz}$ spin down states (with an isosurface
value of 0.001 e/\AA$^3$), i.e. those where the weight of these $d$ orbitals for spin down is largest, calculated with GGA. These states are located at $E-E_\mathrm{F}= 0.29$ eV
($yz$) and 2.29 eV ($xz$) and move across the Fermi level when the atomic charge of the metal increases towards the Zn. Notice that in the case of the $d_{yz}$ state, due to the
stronger hybridization, the $d$ peak splits in two and therefore the choice is ambiguous. The spatial distribution of each peak is similar however. On each $d$ state it
is possible to see the typical shape associated to it, i.e. four lobes on the diagonals of the  $YZ$ or $XZ$ planes, plus some charge on other atoms of the molecule due to
hybridization with other molecular orbitals. Notice that the $d_{yz}$ state does not interact too much with the nitrogens but rather with the carbons, specially with
those located closest to the sulphur atoms. This produces an hybridization between this orbital and the carbon $\pi$ states, as can be clearly seen in
Fig. (\ref{Fe.XZ-YZ}) where the charge on the carbon atoms is mainly located on top of them. On the other hand, in the $d_{xz}$ case, the lobes are directed exactly
towards the nitrogen atoms and therefore this state interacts strongly with them. This interaction is $\sigma$-like, which is the type of interaction that nitrogen forms
with the adjacent atoms, and therefore localizes the charge between atoms. The differences between the $d_{yz}$ and $d_{xz}$ orbitals and their coupling to different
molecular states has a direct impact on the transport properties of some of these molecules, as explained in next section.

\begin{figure}
\includegraphics[width=\columnwidth]{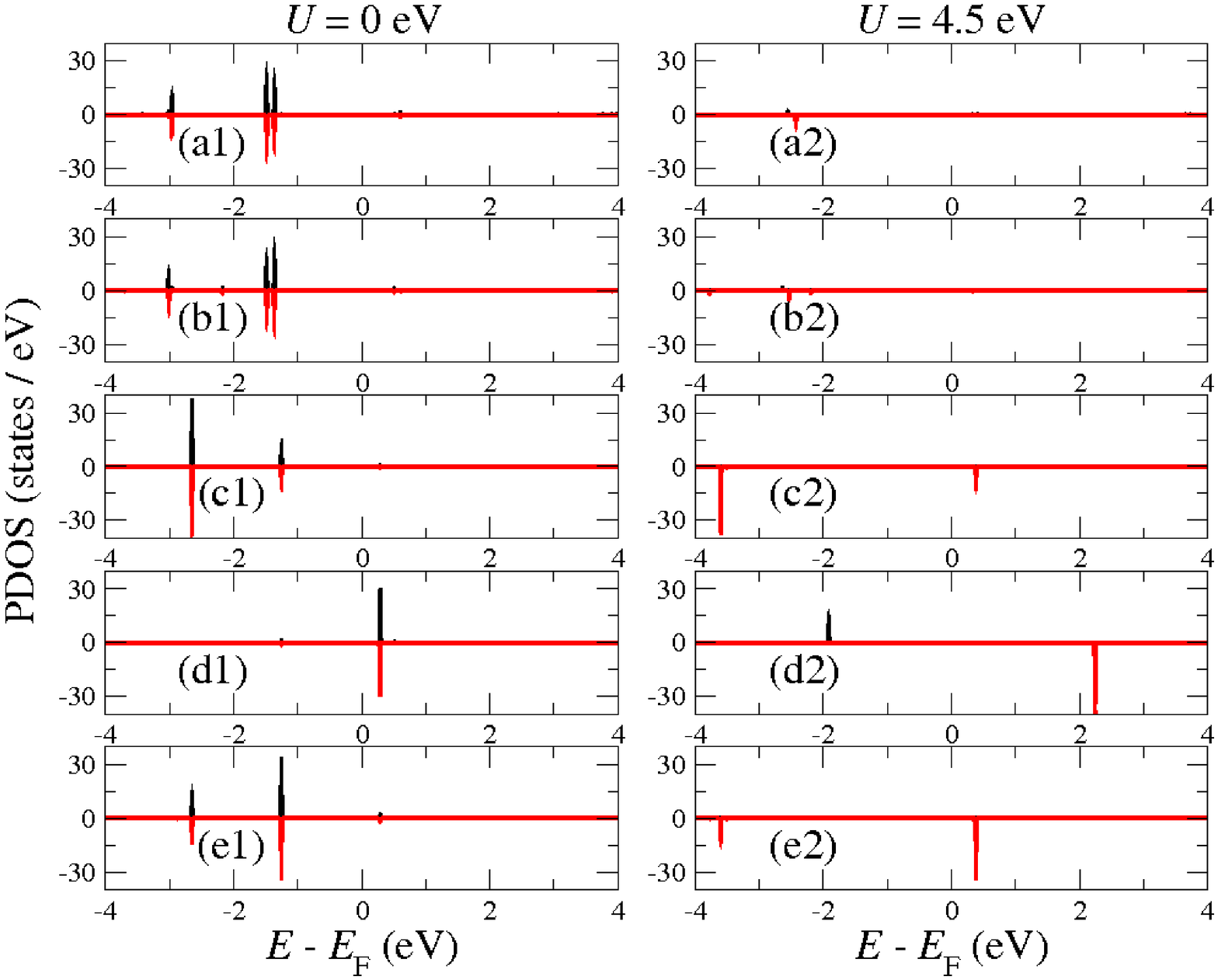}
\caption{\label{PDOS.GGA.SP.Ni}(Color online) Projected density of states (PDOS) on the nickel $d$ states $- d_{xy}$ (a), $d_{yz}$ (b), $d_{z^2}$ (c), $d_{xz}$ (d) and
$d_{{x^2}-{y^2}}$ (e) $-$ for a nickel metalloporphyrin in the gas phase, computed with GGA and spin polarization. Left (1) and right (2) columns correspond to calculations with
$U=0$ eV and $U=4.5$ eV. Positive and negative values represent spin up and spin down electrons.}
\end{figure}

\subsection{CoP}

The lowest energy electronic configuration for CoP, in all cases, is [...]$ {(d_{xy})^2} {(d_{z^2})^2} {(d_{yz})^2} {{(d_{{x^2}-{y^2}}})^1}$, which produces a ground state with
$S_z=1/2$, in agreement with previous results \cite{MengSheng02}. With GGA, the HOMO is made of Fe $3d_{yz}$ and $3d_{xy}$ states,
C $2p_{y}$ states and small fraction of N $2p_{y}$ states. The LUMO is made only of Co $3d_{z^2}$ and a small fraction of $3d_{{x^2}-{y^2}}$. The gap is about 0.9 eV.
With GGA$+U$, in general, all states (Fe and C) lower their energy. Now the HOMO and LUMO are formed by C $2p_{y}$ and N C $2p_{y}$ states and have a small contribution of
Co $3d_{xz}$ states; the GGA $+U$ is 1.8 eV. With LDA the results are similar, but the HOMO-LUMO gap is 0.3 eV, which increases to 1.7 eV for LDA$+U$.

\subsection{NiP}

Unlike the FeP and CoP cases, the ground state electronic configurations of NiP changes when the $U$ is included. For GGA and LDA, the electronic configuration
is [...]$ {(d_{xy})^2} {(d_{z^2})^2} {(d_{yz})^2} {{(d_{{x^2}-{y^2}}})^2}$, i.e. $S_z=0$, which makes the molecule diamagnetic. When the $U$ is introduced,
the electronic configuration becomes [...]$ {(d_{xy})^2} {(d_{z^2})^2} {(d_{yz})^2} {{(d_{{x^2}-{y^2}}})^1} {(d_{xz})^1}$. The spin is $S_z=1$, and the
molecule becomes magnetic. This is produced by the transfer of one electron from the $3d_{{x^2}-{y^2}}$ to the $3d_{xz}$ orbital. This can clearly be seen in
Fig. (\ref{PDOS.GGA.SP.Ni}), where without $U$ both spin up and spin down $3d_{xz}$ states are above the Fermi level (panel (d1)) whereas both $3d_{{x^2}-{y^2}}$ are below
the Fermi level (panel (e1)). However, when $U$ is introduced one of the $3d_{xz}$ moves downwards and one of the $3d_{{x^2}-{y^2}}$ moves upwards, each of them
crossing the Fermi level. It can be therefore concluded that Hund's interaction is enhanced when $U$ is introduced due to the larger repulsion between electrons.

\begin{figure}
\includegraphics[width=\columnwidth]{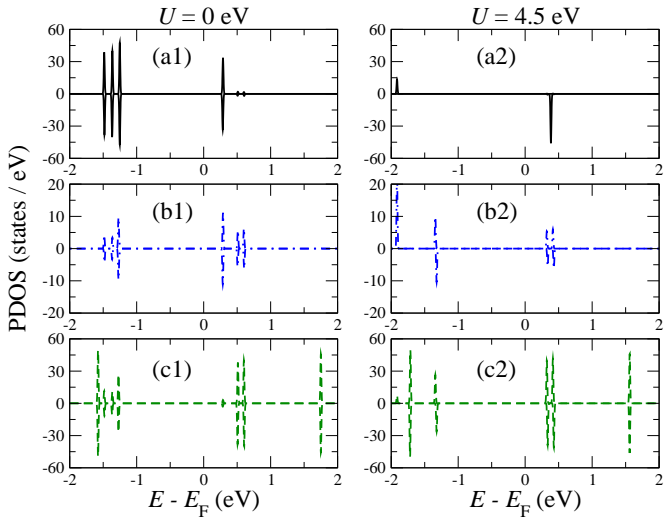}
\caption{\label{PDOS.GGA.SP.NiP}(Color online) Projected density of states (PDOS) on the nickel (a), nitrogen (b) and carbon (c) states for a nickel metalloporphyrin in the gas
phase, computed with GGA and spin polarization. Left (1) and right (2) columns correspond to calculations with $U=0$ eV and $U=4.5$ eV. Positive and negative values represent spin
up and spin down electrons. Notice the vertical scale is different in the middle panels (nitrogen).}
\end{figure}

Fig.(\ref{PDOS.GGA.SP.NiP}) shows the PDOS for Ni, C and N. With GGA the HOMO is formed by Ni, C and N states. Specifically, by Ni $3d_{z^2}$ and $3d_{{x^2}-{y^2}}$ and by
C and N $2p_{y}$ states. Unlike previous cases, the states below the Fermi level with spin up compensate the states with spin dow and the molecule becomes diamagnetic.
The LUMO, is formed by Ni $3d_{xz}$ states and has a small contribution of N $2p_{x}$ and $2p_{z}$ states. The gap is about 1.55 eV. However, when $U$ is introduced,
the HOMO is just made of C and N $2p_{y}$ states. The LUMO is made of Ni $3d_{z^2}$ and $3d_{{x^2}-{y^2}}$ and by C and N $2p_{y}$ states, just like the HOMO with GGA.
In addition, as said before, the molecule becomes magnetic with $S_z=1$. The gap with GGA$+U$ is similar to the case without $U$ (1.65 eV).

Similar words can be said when comparing LDA with LDA$+U$. With LDA the molecule is again diamagnetic, with a gap of 1.55 eV, like GGA. In this case however, the HOMO is
only formed by Ni $3d$ states and in the LUMO the C $2p_{y}$ states have stronger weight. With LDA$+U$, the HOMO is formed by C and N $2p_{y}$ states, and the LUMO by
Ni $3d_{z^2}$ and $3d_{{x^2}-{y^2}}$ and by C and N $2p_{y}$ states, just like with GGA$+U$. The gap in this case does not change (1.55 eV).

\begin{table}
\caption{Change in Mulliken populations, $\Delta e$, when the metalloporphyrins are coupled between gold leads, calculated with LDA and GGA and with or without $U$.}\label{Tab03}
\begin{ruledtabular}
\begin{tabular}{l|lcccc}
&&LDA &LDA$+U$&GGA&GGA$+U$ \\
\hline
{\multirow{5}{*}{Spin up}}
&AuFePAu&0.097&0.194&0.134&0.211\\
&AuCoPAu&0.115&0.260&0.191&0.286\\
&AuNiPAu&0.255&-0.729&0.281&-0.708\\
&AuCuPAu&0.164&0.236&0.181&0.277\\
&AuZnPAu&0.273&0.276&0.301&0.297\\
\hline
{\multirow{5}{*}{Spin down}}
&AuFePAu&0.545&0.360&0.529&0.380\\
&AuCoPAu&0.495&0.246&0.401&0.263\\
&AuNiPAu&0.254&1.269&0.276&1.290\\
&AuCuPAu&0.444&0.315&0.459&0.309\\
&AuZnPAu&0.273&0.276&0.297&0.296\\
\end{tabular}
\end{ruledtabular}
\end{table}

\subsection{CuP}

In CuP, the lowest energy electronic configuration is the same with and without $U$: [...]$ {(d_{xy})^2} {(d_{z^2})^2} {(d_{yz})^2} {{(d_{{x^2}-{y^2}}})^2} {(d_{xz})^{1.5}}$.
In this case, the $3d$ orbitals have closed shells, except the $3d_{xz}$, which loses a bit of charge. The total spin of the molecule is $S_z=1/2$, which mainly comes
from the Cu $4s$ and $3d_{xz}$ orbitals. Furthermore, the magnetism is not strictly localized on the Cu, but extends a little to the nitrogens.

The PDOS on Cu, C and N with GGA, shows that the HOMO comes from C and N $2p_{y}$ states. The Cu states are present in the HOMO-1, which is made of Cu $3d_{xz}$ and has small
contributions from N and C $2p_{x}$ and $2p_{z}$ states. The LUMO is made of Cu $3d_{xz}$ and N $2p_{x}$ states. The energy gap is 0.75 eV. With GGA$+U$, the HOMO remains
the same, but the previous HOMO-1 moves to lower energies and the new HOMO-1 has C and N $2p_{y}$ character. Although the LUMO is still formed by the same type of orbitals
(i.e. Cu $3d_{xz}$ and N $2p_{x}$) now changes its spin polarization and becomes populated by spin down electrons, unlike GGA. The energy gap with $U$ is about 1.4 eV.

With LDA the PDOS is similar to the GGA case, since the nearest states to Fermi level are the spin up Cu $3d_{xz}$ , N and C  $2p_{x}$ and $2p_{z}$ states.
We therefore consider the HOMO to be located, as before, in the carbons and nitrogens. The HOMO-LUMO gap between spin-down states is 0.85 eV, which is now smaller than the
HOMO-LUMO gap between spin-up states, 1.40 eV. The LUMO is the same as with GGA and is populated again with spin-down electrons. The results with LDA$+U$ are the same as
with GGA$+U$, but shifted slightly in energy.

\subsection{ZnP}

The lowest energy electronic structure for ZnP is a close-shell electronic configuration (all the Zn $3d$ and $4s$ states are filled) due to the fact that the Zn states are
all very deep in energy as a consequence of the large nuclear attraction. Therefore the $U$ correction does not affect to electronic configuration of this molecule. By
analyzing the PDOS for Zn, C and N it can be shown that the HOMO and LUMO come from N and C $2p_{y}$ states. The energy gaps in all cases are about 1.65 eV. The only
difference between LDA (LDA$+U$) and GGA (GGA$+U$) is the energy difference between levels, which is slightly different.

\begin{figure}
\includegraphics[width=\columnwidth]{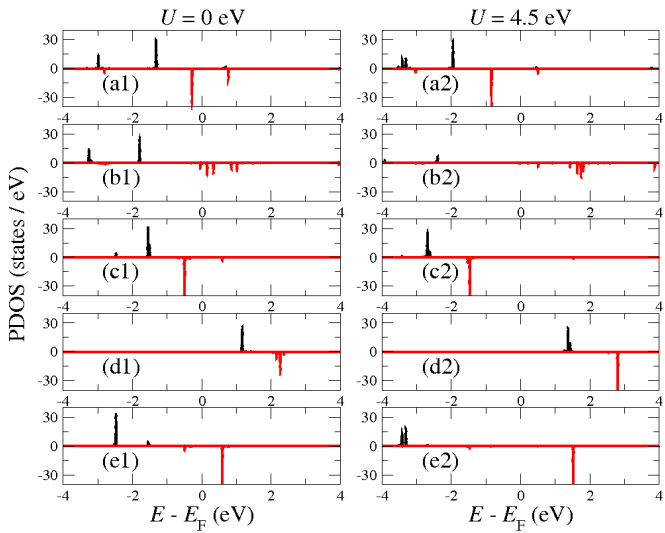}
\caption{\label{PDOS.GGA.SP.Fe-Au}(Color online) Projected density of states (PDOS) on the iron $d$ states $- d_{xy}$ (a), $d_{yz}$ (b), $d_{z^2}$ (c), $d_{xz}$ (d) and $d_{{x^2}-{y^2}}$ (e) $-$ for a iron metalloporphyrin between gold electrodes, computed with GGA and spin polarization. Left (1) and right (2) columns correspond to calculations with $U=0$ eV and $U=4.5$ eV. Positive and negative values represent spin up and spin down electrons.}
\end{figure}

\section{Electronic structure and transport properties of metalloporphyrins between gold electrodes}

\subsection{Electronic structure}

When the metalloporphyrine molecules are coupled to the electrodes the most important effect on the PDOS is the broadening of molecular orbitals into resonances, as a
consequence of the coupling and hybridization of the molecular states with the gold states. Although this effect is not very large on the $d$ states it can still be
seen by comparing Fig. (\ref{PDOS.GGA.SP.Fe}) and Fig. (\ref{PDOS.GGA.SP.Fe-Au}). In case of iron, the largest difference is seen in the $d_{xy}$ PDOS (panel (b)),
which spreads and decreases its height much more. This seems to indicate a larger delocalization of this state as a consequence of its coupling to other molecular states
that hybridize with the gold states. Something similar happens to the $d_{xz}$ and, to a lesser extent, to the other $d$ orbitals. This effect is maintained when the $U$
is included. Also, as a consequence of hybridization and charge transfer the total spin of the molecule is reduced to almost a half of the value of the isolated molecule.
This reduction comes mainly from the $d_{xy}$ state, which spreads and crosses the Fermi level. The charge transfer is summarized in
Table \ref{Tab03}, which shows the change in the Mulliken populations of the molecule when it is coupled between gold leads. As can be seen, all molecules gain charge,
excluding the `pathological' case of Ni. Since the gain is larger for spin up than for spin down the molecule ends up decreasing its magnetic moment, which is consistent
with more delocalized states that reduce Hund's interaction.

\begin{figure}
\includegraphics[width=\columnwidth]{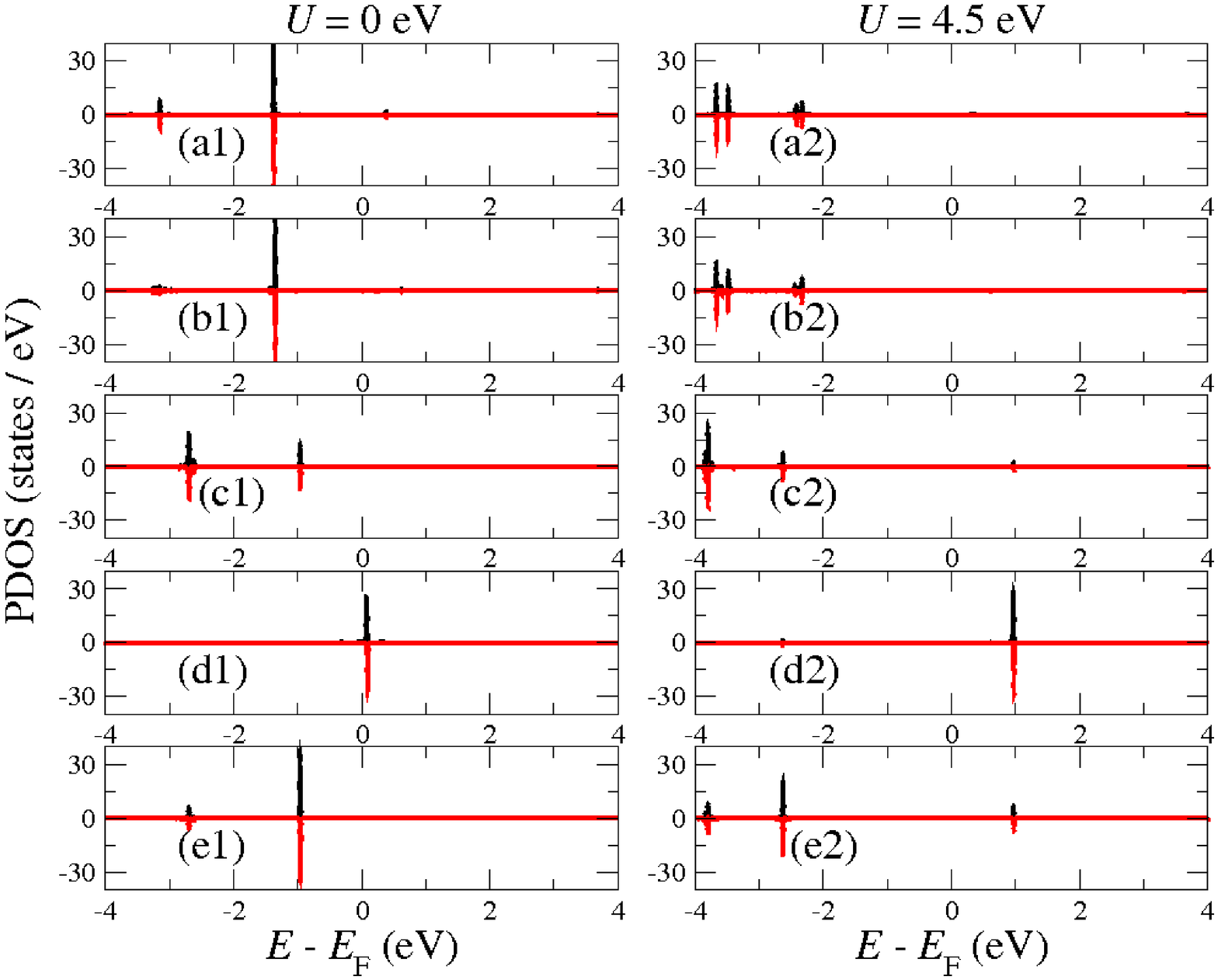}
\caption{\label{PDOS.GGA.SP.Ni-Au}(Color online) Projected density of states (PDOS) on the nickel $d$ states $- d_{xy}$ (a), $d_{yz}$ (b), $d_{z^2}$ (c), $d_{xz}$ (d) and $d_{{x^2}-{y^2}}$ (e) $-$ for a nickel metalloporphyrin between gold electrodes, computed with GGA and spin polarization. Left (1) and right (2) columns correspond to calculations with $U=0$ eV and $U=4.5$ eV. Positive and negative values represent spin up and spin down electrons.}
\end{figure}

In metalloporphyrins other than FeP, the broadening of resonances is not so clear, which can be explained by taking into account that the $d$ states become more
localized when the atomic number increases. In some cases the $d$ states seem to even become more localized as can be seen by comparing panels (a) and (b) in
Fig. (\ref{PDOS.GGA.SP.Ni}) and Fig. (\ref{PDOS.GGA.SP.Ni-Au}). The case of Ni is also the most striking since the differences between the coupled and the isolated molecule
are not only quantitative but qualitative. As can be seen in Fig. (\ref{PDOS.GGA.SP.Ni-Au}), there is no splitting between up and down levels both with and without $U$, as
opposed to the case with $U$ in the isolated molecule. This can be explained by arguing that electrons in the $d_{xz}$ and $d_{{x^2}-{y^2}}$ can be now more
delocalized, which decreases the effect of Hund's interaction. As a consequence, the total magnetic moment with and without $U$ in this case is 0.00 $\mu_\mathrm{B}$, as
can also be seen in Table \ref{Tab02}.

\subsection{Transport properties}

\begin{figure}
\includegraphics[width=\columnwidth]{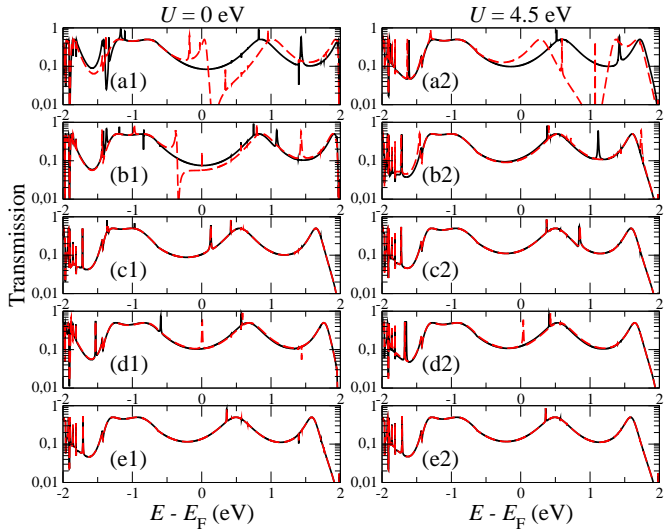}
\caption{\label{TRC.LDA.SP}(Color online) Transmission coefficients for Fe (a), Co (b), Ni (c), Cu (d) and Zn (e) metalloporphyrins between gold electrodes, computed with
LDA and spin polarization. Left (1) and right (2) columns correspond to calculations with $U=0$ eV and $U=4.5$ eV. Continuous and dashed lines represent spin up and
spin down electrons.}
\end{figure}

\begin{figure}
\includegraphics[width=\columnwidth]{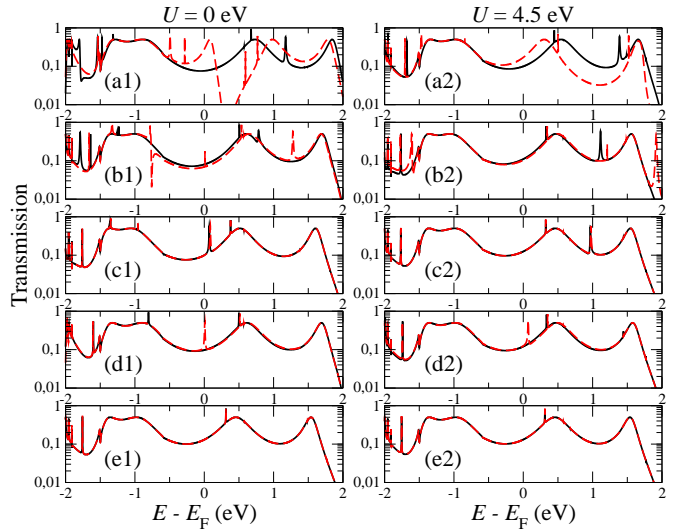}
\caption{\label{TRC.GGA.SP}(Color online) Transmission coefficients for Fe (a), Co (b), Ni (c), Cu (d) and Zn (e) metalloporphyrins between gold electrodes, computed
with GGA and spin polarization. Left (1) and right (2) columns correspond to calculations with $U=0$ eV and $U=4.5$ eV. Continuous and dashed lines represent spin up
and spin down electrons.}
\end{figure}

\begin{figure}
\includegraphics[width=\columnwidth]{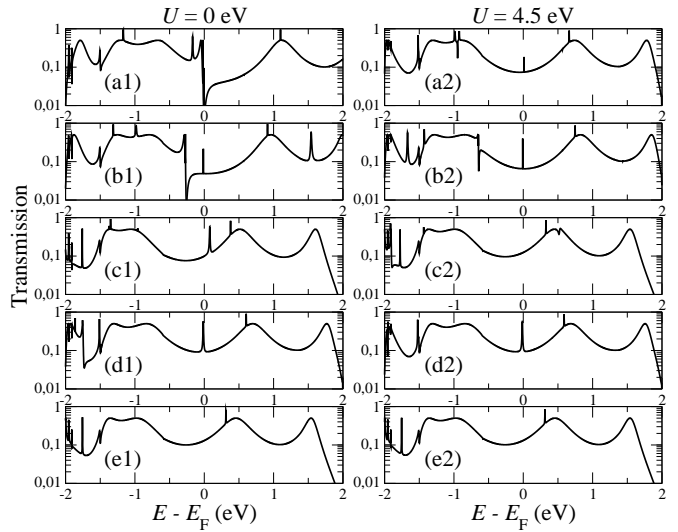}
\caption{\label{TRC.GGA.No_SP}(Color online) Transmission coefficients for Fe (a), Co (b), Ni (c), Cu (d) and Zn (e) metalloporphyrins between gold electrodes,
computed with GGA and without spin polarization. Left (1) and right (2) columns correspond to calculations with $U=0$ eV and $U=4.5$ eV.}
\end{figure}

The zero-bias transport properties of these molecules are shown in Figs. (\ref{TRC.LDA.SP}), (\ref{TRC.GGA.SP}) and (\ref{TRC.GGA.No_SP}), calculated with LDA and
spin polarization, GGA and spin polarization and GGA without spin polarization, respectively. As can be seen, increasing the atomic number from Fe to Zn produces
qualitative changes in those transport properties at the beginning of the series. The main difference between the different metalloporphyrins is due to two types
of resonances that move to lower energies as the atomic number increases. One is a Fano-like resonance, a typical interference effect \cite{Pat97,Ric10,Spa11,Kal12,Gue12},
which is shown as a sharp increase followed by a steep decrease of the transmission. The other seems to be a sharp Breit-Wigner resonance. These resonances appear in
all cases, specially for Fe and Co, with GGA or LDA, with or without $U$ and with or without spin polarization. The only differences are their position and width.
With LDA and spin polarization, Fig. (\ref{TRC.LDA.SP}), the Fano-like resonance appears clearly with Fe and, much sharper and on lower energies, with Co. This resonance
moves to higher (lower) energies when $U$ is added to Fe (Co). Above this resonance there is a Breit-Wigner-like resonance that moves to lower energies from Fe to Zn and
appears just at the Fermi level on CuP. It moves also to higher or lower energies when the $U$ is included. With LDA and spin polarization, Fig. (\ref{TRC.GGA.SP}),
the results are similar. Without spin polarization, Fig. (\ref{TRC.GGA.No_SP}) we show for simplicity just the GGA cases. Again the behavior is similar. Some of these
last results without $U$ are also analogous to those obtained by Wang {\em et al.} \cite{Wan09}

\begin{figure}
\includegraphics[width=\columnwidth]{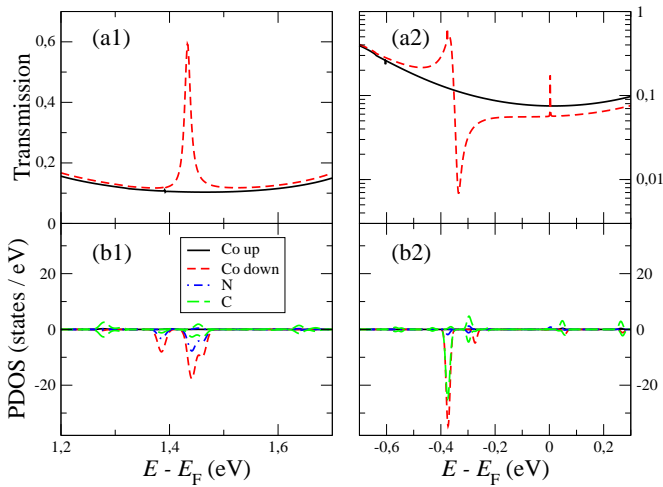}
\caption{\label{TRC-PDOS.BW-Fano}(Color online) Transmission coefficients (a) and PDOS (b) around a Breit-Wigner-like resonance (1) and a Fano resonance (2), calculated
on cobalt metalloporphyrin between gold electrodes and computed with LDA and spin polarization. Positive and negative values in the lower panels represent spin up
and spin down electrons.}
\end{figure}

Obviously the two types of resonances must come from the $d$ orbitals, due to their evolution with the metallic atom and also the fact that they are not present in
porphyrins without metallic atoms. One hint about their origin can be obtained by looking at the projected density of states of the $d$ levels and the surrounding atoms
where these resonances happen. This is shown in Fig. (\ref{TRC-PDOS.BW-Fano}) for the case of the Cobalt metalloporphyrin, computed with LDA and without $U$, which is
the configuration where different contributions from different atoms cam be most clearly seen. First, the spin down $d$ states that are closest to the energy of the resonance
are the $d_{xz}$ and $d_{yz}$ in the Breit-Wigner-like and Fano-like cases, respectively. For the first resonance the PDOS shows that there is a rather strong
hybridization between the $d_{xz}$ level and the nitrogen atoms. For the Fano-like resonance, however, the $d_{yz}$ hybridizes much more with the carbon atoms.
This is in agreement with the LDOS of Fig. (\ref{Fe.XZ-YZ}), which shows that the $d_{xz}$ hybridizes with the nitrogens and the $d_{yz}$ with the carbons.

Taking into account the previous data it is possible to explain the behavior of these systems as follows. The $d$ orbitals are localized states that couple to certain
molecular states. Such configuration is similar to that where a side state couples to a molecular backbone, which produces Fano resonances in the transmission coefficients.
The $d$ orbitals generate therefore Fano resonances associated to one or various molecular states. If, for example, the molecular state is the LUMO and the $d$ orbital is in
the HOMO-LUMO gap, the effect of the Fano resonance is clearly seen because it affects the transmission in the gap, which has a large weight from the LUMO. The Fano resonance
does not go to zero, however, because of the tails of the transmission of other molecular states. If, on the other hand, the $d$ orbital couples to a state below the HOMO or
above the LUMO, the effect of the Fano resonance turns out to be much smaller because the transmission of such state does not affect the transmission in the HOMO-LUMO gap.
There is however one effect produced by the Fano resonance, which is related to its peak. This peak has a transmission of 1 at the tip and therefore can be seen above
the background of the transmission of other states. We can therefore confirm that the sharp Breit-Wigner-like resonances that are seen at different energies come in reality
from Fano resonances. In view of these results it is possible to conclude then that the Fano features inside the HOMO-LUMO gap are produced by $d$ orbitals that couple to
the $\pi$ states, appear mainly on the LUMO. One $d$ orbital responsible for such a feature is clearly the $d_{xy}$, as seen in Fig. (\ref{Fe.XZ-YZ}). On the other hand, the
sharp resonances are generated by $d$ orbitals, such as the $d_{xz}$, that couple to sigma states above the LUMO or below the HOMO.

These features, specially the sharp resonance, appear in all cases and are a general characteristic of these molecules. We propose that the above-discussed metalloporphyrine
molecules could be used as atomic or gas sensors. This is so due to the sharp changes in the zero-bias conductance that they generate when the Fano resonances cross the
Fermi level, which should be altered whenever a given atom or molecules couples either to the metallic atom or to other parts of the molecular backbone. These
resonances induce also differences between the spin-up and spin-down transmissions, which are specially important in the case of iron, cobalt and copper. These could give
rise to spin-filtering properties. The inclusion of the $U$ term reduces however the spin-filtering behavior, which only remains in the case of FeP.

\begin{figure}
\includegraphics[width=\columnwidth]{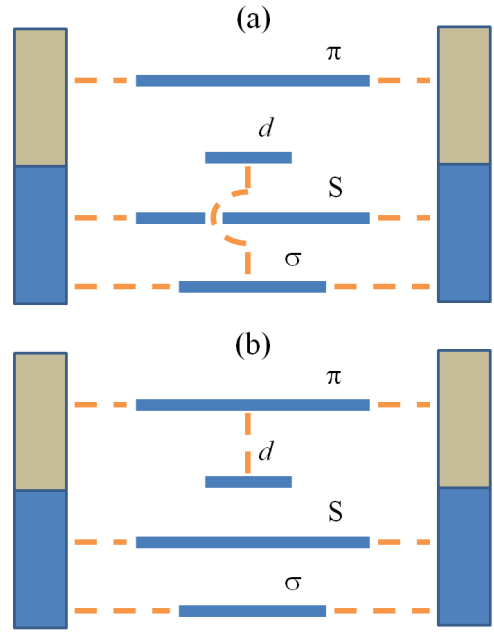}
\caption{\label{Model}(Color online) Schematic representation of the model used to reproduce the ab-initio results. Case (a) corresponds to the $d$ level coupled to
the $\sigma$ state (labeled in the text as $H_1$) and case (b) to the $d$ level coupled to one of the $\pi$ levels ($H_2$).}
\end{figure}

\subsection{Simple model}

The previous behavior can be described with a simple model that takes into account the coupling of the molecular orbitals placed in the neighborhood of the Fermi
energy to featureless leads displaying a flat density of states, as well as the coupling of one of those molecular orbitals to the $d$ orbital responsible for the Fano
resonance\cite{Mar09,Gar11}. Specifically, the model includes four molecular levels: a $\sigma$ level below the HOMO, a level associated to the two linker sulphur
atoms that represents the HOMO and a $\pi$ level which represents the LUMO. Finally, the $d$ level has no direct coupling to the electrodes, but is instead coupled
either to the $\sigma$ or to the $\pi$ levels. The model and its two possible couplings are sketched
in Fig. (\ref{Model}) (a) and (b), respectively. The Hamiltonian is therefore diagonal, except for the coupling between a given state and the $d$ level. We also
choose diagonal and identical $\Gamma$ matrices, that couple the molecule to the leads:

\begin{eqnarray}
 \hat H&=&\left(\begin{array}{cccc}
\epsilon_\sigma&0&t_\sigma&0\\
0&\epsilon_S&0&0\\
t_\sigma^*&0&\epsilon_d&t_\pi\\
0&0&t_\pi^*&\epsilon_\pi \end{array}\right),\nonumber\\
\Gamma_L=\Gamma_R&=&\left(\begin{array}{cccc}
\Gamma_\sigma&0&0&0\\
0&\Gamma_S&0&0\\
0&0&\Gamma_d&0\\
0&0&0&\Gamma_\pi \end{array}\right)
\end{eqnarray}

\noindent To facilitate the comparison with the ab-initio results, we have chosen the following values for the on-site energies and the couplings: $\epsilon_\sigma=-2$,
$\epsilon_S=-1$, $\epsilon_d=0$, $\epsilon_\pi=0.6$, $\Gamma_\sigma=0.04$, $\Gamma_S=0.06$, $\Gamma_d=0$ and $\Gamma_\pi=0.06$. where all energies are measured in $eV$.
For model (a), we choose $t_\sigma=-0.4$ and $t_\pi=0$, while for model (b) we choose $t_\sigma=0$ and $t_\pi=-0.2$. The transmission and the retarded Green's
function are then given by

\begin{eqnarray}
 T(E)&=&\mathrm{Tr}\left[\hat\Gamma\hat G^{\mathrm{R}\dag}(E) \hat\Gamma\hat G^\mathrm{R}(E)\right]\nonumber\\
 \hat G^\mathrm{R}(E)&=&\left[E\hat I-\hat H_{a,b}- i\hat\Gamma\right]^{-1}
\end{eqnarray}

\begin{figure}
\includegraphics[width=0.8\columnwidth]{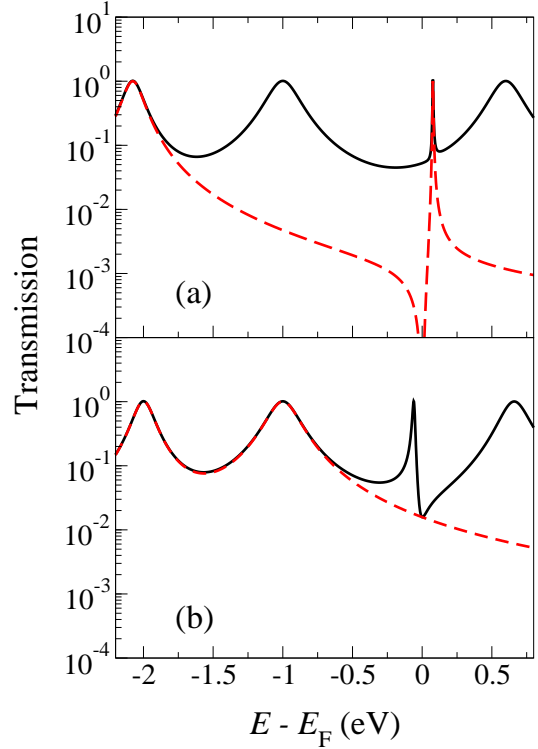}
\caption{\label{TRC.Model}(Color online) (a) Transmission coefficients calculated for model (a) described in the text and in Fig. (\ref{Model}) (a). The black solid line corresponds to
the full model (a), whereas the red dashed corresponds to a simplified (a) model, where the $S$ and $\pi$ orbitals have been dropped from the calculation; (b) Transmission
coefficients calculated for model (b) described in the text and in Fig. (\ref{Model}) (b). The black solid line corresponds to the full model (b), whereas the red dashed corresponds to a simplified (b) model, where the $S$ and $\sigma$ orbitals have been dropped from the calculation;}
\end{figure}

We show in Fig. (\ref{TRC.Model}) the transmissions obtained for models (a) and (b). The dashed line in Fig. (\ref{TRC.Model}) (a) shows the transmission coefficient found when the sulfur and
$\pi$ orbitals in model (a) are dropped and only the d and $\sigma$ orbitals are  considered. A clear-cut Fano resonance emerges at the d level on-site energy. However, this
resonance is masked when the the sulfur and $\pi$ orbitals are re-integrated back into the calculation leaving what looks at first sight a conventional Breit-Wigner resonance.
the dashed line in Fig. (\ref{TRC.Model}) (b) is the transmission obtained in model (b) when the $\sigma$ and sulfur orbitals are left aside, which features again a Fano resonance. However,
the Fano resonance is now clearly visible in the full model, because the position of the $\sigma$ and sulfur resonance combined with the ordering of the divergencies in the
Fano resonance can not mask completely the drop in transmission at higher energies. The similarity with the ab-initio results for the cobalt porphyrine shown in Fig.
(\ref{TRC.LDA.SP}) is striking, despite the simplicity of the model.

\section{Conclusions}

We have studied the electronic properties of isolated metalloporphyrins and determined the influence of the exchange-correlation potential and strong correlations.
We found that the HOMO-LUMO gap is greatly improved when a LDA+$U$ or GGA+$U$ description is used. However, we have found that the spin of the molecule does not change,
excluding the case of NiP. We have also studied the electronic and transport properties of metalloporphyrins between gold electrodes. We found that the coupling to the
electrodes changes only slightly the electronic properties but the magnetic moments decrease as a consequence of charge transfer and hybridization with the electrodes.
We have found two types of features in the transport properties that we show to be Fano resonances by the use of a simple model. We propose the use of metalloporphyrine
molecules as possible nanoelectronic devices with sensing and spin-filtering functionalities.

\acknowledgments This work was supported by the the Spanish Ministry
of Education and Science (project FIS2009-07081) and the Marie Curie
network NanoCTM. VMGS thanks the Spanish Ministerio de Ciencia e Innovaci\'on for a Ram\'on y Cajal fellowship (RYC-2010-06053). RF acknowledges financial support through grant Severo-Ochoa (Consejer\'{\i}a de Educaci\'on, Principado de Asturias). We acknowledge discussions with C. J. Lambert.

{}

\end {document}